\documentclass[twocolumn,amsmath,amssymb,pre]{revtex4}

\usepackage{graphicx}
\usepackage{bm}

\usepackage{amsfonts}
\usepackage{amsmath,amsfonts,amsthm,bm}

\usepackage[utf8]{inputenc}

\begin{document}
\title{Dynamics of multidimensional fundamental and vortex solitons in random media}

\author{Volodymyr M. Lashkin}
\email{vlashkin62@gmail.com} \affiliation{$^1$Institute for
Nuclear Research, Pr. Nauki 47, Kyiv 03028, Ukraine}
\affiliation{$^2$Space Research Institute, Pr. Glushkova 40 k.4/1,
Kyiv 03187,  Ukraine}

\begin{abstract}
We study the dynamics of fundamental and vortex solitons in the
framework of the nonlinear Schr\"{o}dinger equation with the
spatial dimension $D\geqslant 2$ with a multiplicative random term
depending on the time and space coordinates. To this end, we
develop a new technique for calculating the even moments of the
$N$th order. The proposed formalism does not use closure
procedures for the nonlinear term, as well as the smallness of the
random term and the use of perturbation theory. The essential
point is the quadratic form of the autocorrelation function of the
random field and the special stochastic change of variables. Using
variational analysis to determine the field of structures in the
deterministic case, we analytically calculate a number of
statistical characteristics describing the dynamics of fundamental
and vortex solitons in random medium, such as the mean
intensities, the variance of the intensity, the centroid and
spread of the structures, the spatial mutual coherence function
etc. In particular, we show that, under the irreversible action of
fluctuations, the solitons spread out, i.e., no collapse occurs.
\end{abstract}

\maketitle

\section{Introduction }

The propagation of nonlinear waves in random media is an important
problem in nonlinear physics and has been intensively studied over
the past few decades, i. e. almost immediately with the widespread
use of the concept of soliton in physics
\cite{Bass1988,Gredeskul1992,Konotop_book1994,Abdullaev_Garnier2005}.
This is due both to general theoretical interest, where the
problems of fundamental nonlinear (primarily soliton) physics and
statistical physics are intertwined, but also from a practical
point of view of applied problems of optics, acoustics, plasma
physics, physics of disordered systems, etc., where fluctuating
parameters and randomness appears in a natural way. Depending on
the problems under study, the source of fluctuations can be the
refractive index of the medium, which changes randomly and
continuously in time and space, the presence of turbulence in
laboratory and space plasmas and in the ionosphere and
troposphere, particles randomly distributed in space (e. g. rain,
fog, hail), random interaction potential in disordered systems,
etc. \cite{Tatarski_book1967,Ishimaru_book,Gredeskul1988}. Note
that, as is well known, the wave equation in a random medium even
in the linear case cannot be solved explicitly in the sense that
it is impossible to determine the probability distribution of the
wave field as a function of the probability distribution of the
medium \cite{Tatarski_book1967,Ishimaru_book}. This applies even
more to the propagation of nonlinear waves in random media. The
fundamental difficulty faced by any statistical description of
nonlinear waves is the statistical closure problem, in which the
time evolution of the moments of order $n$ for the wave field is
coupled by the nonlinearity to moments of order $n+1$. To obtain
closed nonlinear equations for statistical characteristics (for
example, moments of the $n$th order), various approximate
statistical closure procedures are used. This provides a way of
expressing a moment of some order in terms of lower-order moments.
Nevertheless, a fairly extensive literature is devoted to
nonlinear waves in random media. In the vast majority of cases,
this concerns the one-dimensional (1D) case that is soliton
propagation in random media. In particular, the propagation of
solitons in the presence of randomness in models of the nonlinear
Schr\"{o}dinger (NLS) equation, the Korteweg-de Vries (KdV)
equation and the sine-Gordon equation, as well as shock waves of
the stochastic Burgers equation, has been studied in sufficient
detail
\cite{Abdullaev1982,Elgin1985,Gordon-Haus1986,Malomed1987,Bass-Kivshar1988,
Sanchez1990,Garnier2001,Abdulaev2004,Conti2012,Fishman2012,Schwiete2013,Weiss2016}
(see also the reviews and books
\cite{Bass1988,Gredeskul1992,Konotop_book1994,Abdullaev_Garnier2005,Kivshar-Malomed1989}
and references therein). The phenomenon of Anderson localization
in disordered nonlinear systems with a random potential within the
framework of the discrete NLS equation model was studied in
Refs.~\cite{Pikovsky2008,Flach2009,Milovanov2012}, where the
unlimited spreading of an initially localized wave packet was
shown. Attempts were also made to directly numerically simulate
solitons within the framework of the stochastic KdV and NLS
equations \cite{Debussche1999,Debussche2002}. Examples of the 1D
solitons in random media with a nonlocal nonlinearity were
considered in Refs.~\cite{Molli2010,Skupin2012}. Alfv\'{e}n and
ion-cyclotron solitons in plasma with fluctuating parameters in
the models of the derivative NLS equation and the Fokas-Lenells
equation were studied in
Refs.~\cite{Ruderman2002,Lashkin_Alfen2006} and
\cite{Lashkin2021}, respectively. Solitons of the modified NLS
equation in the presence of a random addition to the refractive
index were considered in
Refs.~\cite{Doktorov2001,Lashkin_MNLS2004}. Two-dimensional (2D)
solitons of the Kadomtsev-Petviashvili equation with a noise term
were explored in Ref.~\cite{Gorodtsov2000}. Much less attention
has been devoted to multidimensional solitons and nonlinear
structures in random media. The point is that, as is known,
stationary solutions (if any) of nonlinear evolution equations in
the multidimensional case often turn out to be unstable and
subject to the phenomenon of collapse or self-focusing
\cite{Chiao1964,Kelley1965,Zakharov1972}. The collapse induced by
the cubic nonlinearity is critical in the 2D space, which means
that it sets in when the norm of the underlying wave field exceeds
a certain finite critical value, and supercritical in the
three-dimensional (3D) space, where an initial state with an
arbitrarily small norm may blow up due to the collapse
\cite{Rubenchik1986,Zakharov1972,Zakharov_UFN2012,Fibich_book,Berge1998}.
Note that to deal with realistic models, saturation had been
suggested as a way to stabilize such a catastrophic collapse or
self-focusing and produce stable soliton structures of higher
dimensions.

Already in early studies
\cite{Pasmanik1974,Jokipii1973,Vorobev1970} on the self-focusing
of optical beams in random media, it was shown that a random
addition to the refractive index can significantly weaken or
completely prevent the effect of self-focusing of the beam. In
this case, different approximations were used, for example, an
incoherent self-interacting light beam was studied in the random
phase approximation in Ref. \cite{Pasmanik1974}, the Rytov method
for wave propagation in random media generalized to include
self-focusing effect was considered in Ref. \cite{Jokipii1973},
and in Ref. \cite{Vorobev1970} a purely Markov approximation was
used. Using the so-called method of polynomial representation of
the moments, it was shown in Ref. \cite{Talanov1971} that, under
fairly general assumptions, collapse does not occur on average. In
Ref. \cite{Gaididei1998}, the results of Ref. \cite{Talanov1971}
were confirmed (in particular, that collapse on average is
impossible) and a sufficient criterion  for arresting the collapse
of a pulse in a nonlinear waveguide with a randomly fluctuating
refractive index was obtained. Note that in all the
above-mentioned works, in the absence of randomness (that is, in
the deterministic case), the considered models were reduced to the
2D NLS equation. The stochastic 2D NLS equation with
multiplicative noise was also considered in
Ref.~\cite{Shivamoggi2000}, where the perturbation theory was
developed in the framework of the so-called
$\varepsilon$-expansion, and in Ref.~\cite{Gorder2010} using the
so-called $\delta$-expansion. In both cases, the authors argued
that the developed perturbation theory makes it possible to
effectively reduce the stochastic 2D NLS equation to the linear
stochastic equation. The effects of randomness have been discussed
in recent papers \cite{Hafizi2017,Penano2017,Deng2020}, and it has
also been noted that stochasticity in multidimensional NLS
equation leads to the prevention of the collapse phenomenon. As
indicated above, the collapse in a multidimensional NLS with a
random term may not arrest in the general case, but be
significantly weakened. First of all, this concerns 3D models
where there is no critical collapse regime. In addition, the
possibility of preventing collapse apparently depends on the type
of random perturbation and its statistical properties. For
example, in Ref.~\cite{Garnier2004} the possibility of collapse in
the 3D NLS model with a random term in the form of noise depending
only on time against the background of a trapping parabolic
potential was demonstrated both by an analytical variational
method and by numerical simulation. It was shown that the expected
collapse time is inversely proportional to the integrated
covariance of the time autocorrelation function.

In most cases, when calculating statistical characteristics the
random term is small and then a perturbation theory along with the
appropriate closure procedures can be applied. One of the few
exceptions is the work of Besieris \cite{Besieris_main} (a brief
synopsis of this work can be found, for example, in
Refs.~\cite{Bass1988,Konotop_book1994}), where a novel statistical
technique is proposed that allows one to calculate the moments of
even orders associated with the 1D NLS stochastic equation without
the need to derive a closed equation for the moments. A necessary
condition for this was the quadratic form of the noise
autocorrelation function. In this work, some statistical
characteristics describing the soliton propagation in the presence
of noise were also calculated.

In the present paper, following the ideas of
Ref.~\cite{Besieris_main}, we consider the NLS equation with
multiplicative random term depending on time and space variables
in the spatial dimension $D\geqslant 2$. In this case, no
procedure of statistical closure for the nonlinear term. The main
thrust of the research will be concerned with the analysis of two
specific stochastic nonlinear equations for even moments employing
the quadratic form of the autocorrelation function. After
introducing a stochastic change of variables and a fundamental
ansatz, we will show explicitly how each of the aforementioned
stochastic nonlinear equations can be mapped into an explicit
deterministic equation for a new dependent variable and a system
of stochastic nonlinear coupled ordinary differential equations
governing the remaining elements of the transformation. The
ensuing analysis proceeds two-fold. At one level, we must be able
to solve a deterministic nonlinear partial differential equation
governing the new dependent variable. At the other level, we must
proceed to derive the Fokker-Planck equation and to find the joint
probability density function for the stochastic system. Combining
the two solutions and inverting the mapping, we are able to find
explicit expressions for any even moment of the wave field. As an
application of the developed formalism, we consider the influence
of noise on the 2D  and 3D fundamental (ground state) and vortex
solitons, which are unstable in the absence of randomness. The
vortex soliton (spinning soliton) is the localized nonlinear
structure with embedded vorticity and ringlike in the 2D or
toroidal in 3D case field intensity distribution, with the dark
hole at the center where the phase dislocation takes place: a
phase circulation around the azimuthal axis is equal to $2m\pi$.
An integer $m$ is referred to as topological charge. We calculate
some averaged observables for these structures and show that both
fundamental and vortex solitons do not collapse.

The paper is organized as follows. In Sec.~II, using
Donsker-Furutsu-Novikov functional formalism, a stochastic
equation for even moments is obtained . The effective nonlinear
stochastic equation and a stochastic change of variables are
proposed in Sec.~III. In Sec.~IV., the Fokker-Planck equation for
the joint probability density is derived and solved. Using the
variational method, analytical expressions for the fundamental and
vortex solitons are found Sec.~V. In Sec.~VI, the statistical
characteristics for the fundamental and vortex solitons are
analytically calculated. The conclusion is made in Sec. VII.

\section{Basic equations}

We start with the normalized NLS equation in the spatial dimension
$D\geqslant 2$ with a multiplicative random term,
\begin{equation}
\label{main} i\frac{\partial\varphi}{\partial
t}+\Delta\varphi+|\varphi|^{2}\varphi+\varepsilon
(\mathbf{r},t)\varphi=0,
\end{equation}
where $\varepsilon (\mathbf{r},t)$ is a zero-mean real homogeneous
in space and stationary in time gaussian random field with the
autocorrelation function
\begin{equation}
\label{correlator1}
\langle\varepsilon(\mathbf{r},t)\varepsilon(\mathbf{r}^{\prime},t^{\prime})\rangle
=A(\mathbf{r}-\mathbf{r}^{\prime})\,C(t-t^{\prime})
\end{equation}
where the angular brackets $\langle \dots \rangle$ means ensemble
averaging. Without loss of generality, we consider both 2D and 3D
cases (the 1D case was considered in Ref. \cite{Besieris_main}).
This model equation arises in a number of important physical
applications. In particular, these are the following.
\begin{enumerate}
\item
The propagation of a continuous wave beam (the 2D case) or the
transmission of a spatiotemporal light signal (the 3D case) in a
bulk self-focusing optical medium with the Kerr nonlinearity and a
random addition to the refractive index. Here $t$ is the
propagation distance $z$, and the $x$- and $y$-axes are two
transverse directions for the 2D case. In the 3D case, the third
coordinate $z$ replaced by the local time $\tau =t-z/v_{0}$
($v_{0}$ is the group velocity of the carrier wave)
\cite{Kivshar_book2003}.
\item The propagation of strong electromagnetic waves in the
turbulent ionosphere \cite{Gurevich_book1978}, where $\varepsilon
(\mathbf{r},t)$ stands for the turbulent part of the refractive
index.
\item The nonlinear Langmuir waves in the turbulent unmagnetized
plasma. Here $\varepsilon (\mathbf{r},t)$ corresponds to the
turbulent part of the plasma density perturbation
\cite{Goldman1984}.
\item Nonlinear matter waves in the Bose-Einstein condensates (BEC) with random
potentials (BEC with disordering) \cite{Yukalov2007,Yu2017}.
\end{enumerate}
Our ultimate goal in connection with Eq. (\ref{main}) is to
compute the $N$th order moments
$\langle\varphi^{(N)}(\mathbf{x},t)\rangle$, where
\begin{equation}
\label{F}
\varphi^{(N)}(\mathbf{x},t)=\prod_{m=1}^{N/2}\varphi(\mathbf{r}_{2m-1},t)
\varphi^{\ast}(\mathbf{r}_{2m},t),
\end{equation}
and
$\mathbf{x}=(\mathbf{r}_{1},\mathbf{r}_{2},\dots,\mathbf{r}_{N})$.
An evolution equation for $\varphi^{(N)}(\mathbf{x},t)$ follows
readily from the definition (\ref{F}) and the basic equation
(\ref{main}) and has the form
\begin{gather}
i\frac{\partial\varphi^{(N)}}{\partial
t}+\sum_{m=1}^{N}\xi_{m}\Delta_{m}\varphi^{(N)}+\sum_{m=1}^{N}\xi_{m}|\varphi(\mathbf{r}_{m},t)|^{2}\varphi^{(N)}
\nonumber \\
+\sum_{m=1}^{N}\xi_{m}\varepsilon(\mathbf{r}_{m},t)\varphi^{(N)}=0,
\label{FN}
\end{gather}
where
\begin{equation}
\Delta_{m}=\frac{\partial^{2}}{\partial
x^{2}_{m}}+\frac{\partial^{2}}{\partial y^{2}_{m}},
\end{equation}
for the two-dimensional case, and
\begin{equation}
\Delta_{m}=\frac{\partial^{2}}{\partial
x^{2}_{m}}+\frac{\partial^{2}}{\partial y^{2}_{m}}+
\frac{\partial^{2}}{\partial z^{2}_{m}}
\end{equation}
for the three-dimensional one. The quantity $\xi_{m}$ entering
into Eq. (\ref{FN}) is assumed to be $\xi_{m}=1$ if $m$ is odd,
and $\xi_{m}=-1$ if $m$ is even. Ensemble averaging of the
equation (\ref{FN}) results in the expression
\begin{gather}
i\frac{\partial\langle\varphi^{(N)}\rangle}{\partial
t}+\sum_{m=1}^{N}\xi_{m}\Delta_{m}\langle\varphi^{(N)}\rangle
+\sum_{m=1}^{N}\xi_{m}\langle|\varphi(\mathbf{r}_{m},t)|^{2}\varphi^{(N)}\rangle
\nonumber \\
+\sum_{m=1}^{N}\xi_{m}\langle\varepsilon(\mathbf{r}_{m},t)\varphi^{(N)}\rangle=0
\label{FN_average}
\end{gather}
with the initial condition
$\langle\varphi^{(N)}(\mathbf{x},0)\rangle=\varphi^{(N)}_{0}(\mathbf{x})$.
Equation (\ref{FN_average}) is not a closed equation for
$\langle\varphi^{(N)}\rangle$ due to the third and fourth terms on
the left-hand side. To calculate an average of the product of the
gaussian random function $\eta(s)$ with the functional $R[\eta]$
depending on it, we will use the Donsker-Furutsu-Novikov
functional formalism \cite{Furutsu1963,Donsker1964,Novikov1965},
where $s$ denotes the set of all arguments, both continuous and
discrete, on which the random function depends. The corresponding
expression is
\begin{equation}
\label{Furutsu} \langle\eta(s)R[\eta]\rangle=\int
\langle\eta(s)\eta(s^{\prime})\rangle\left\langle\frac{\delta
R[\eta]}{\delta \eta (s^{\prime})}\right\rangle ds^{\prime},
\end{equation}
where $\delta/\delta \eta$ denotes the functional derivative, and
the integral extends over the region in which the function is
defined and, as usual, summation over the repeated indices of the
discrete variables is assumed. A partial closure in Eq.
(\ref{FN_average}) can be performed for a special class of random
fields $\varepsilon (\mathbf{r},t)$ when $C(t-t^{\prime})=\delta
(t-t^{\prime})$ in Eq. (\ref{correlator1}), where $\delta (\tau)$
is the Dirac delta function (the white noise in time). Then, using
Eq. (\ref{Furutsu}) we have for the average in the last term of
Eq. (\ref{FN_average}),
\begin{gather}
\left\langle
\varepsilon(\mathbf{r}_{m},t)\varphi^{(N)}(\mathbf{x},t)
\right\rangle=\int
\langle\varepsilon(\mathbf{r}_{m},t)\varepsilon(\mathbf{r}_{m}^{\prime},t^{\prime})\rangle
\nonumber \\
\times\left\langle\frac{\delta \varphi^{(N)}(\mathbf{x},t)}{\delta
\varepsilon(\mathbf{r}_{m}^{\prime},t^{\prime})}\right\rangle
d\mathbf{r}_{m}^{\prime}dt^{\prime} =\int
A(\mathbf{r}_{m}-\mathbf{r}_{m}^{\prime})
\nonumber \\
\times\left\langle\frac{\delta \varphi^{(N)}(\mathbf{x},t)}{\delta
\varepsilon(\mathbf{r}_{m}^{\prime},t)}\right\rangle
d\mathbf{r}_{m}^{\prime}. \label{Furutsu1}
\end{gather}
The functional derivative $\delta \varphi^{(N)}/\delta
\varepsilon$ required in Eq. (\ref{Furutsu1}) can be found from
the original equation (\ref{FN}). To this end we rewrite Eq.
(\ref{FN}) in the integral form
\begin{gather}
\varphi^{(N)}(\mathbf{x},t)=\varphi^{(N)}_{0}(\mathbf{x})+i\int_{0}^{t}
\sum_{m=1}^{N}\xi_{m}[\Delta_{m}+|\varphi(\mathbf{r}_{m},t^{\prime})|^{2}
\nonumber \\
+\varepsilon(\mathbf{r}_{m},t^{\prime})]\varphi^{(N)}(\mathbf{x},t^{\prime})dt^{\prime}.
\label{FN_integral}
\end{gather}
Then, taking the functional derivative in (\ref{FN_integral}) and
using
\begin{equation}
\frac{\delta \varepsilon (\mathbf{r},t)}{\delta \varepsilon
(\mathbf{r}^{\prime},t^{\prime})}=\delta
(\mathbf{r}-\mathbf{r}^{\prime})\delta (t-t^{\prime}),
\end{equation}
(see, e. g. \cite{Konotop_book1994}), we find for $0<t^{\prime}<t$
\begin{gather}
\frac{\delta\varphi^{(N)}(\mathbf{x},t)}{\delta\varepsilon(\mathbf{r}_{m}^{\prime},t^{\prime})}=i
\sum_{n=1}^{N}\xi_{n}\int_{t^{\prime}}^{t}\left\{[\Delta_{n}+\varepsilon(\mathbf{r}_{n},t^{\prime\prime})]\frac{\delta\varphi^{(N)}
(\mathbf{x},t^{\prime\prime})}{\delta\varepsilon(\mathbf{r}_{m}^{\prime},t^{\prime})}\right.
\nonumber \\
\left. +\frac{\delta [|\varphi^{(N)}
(\mathbf{r}_{n},t^{\prime\prime})|^{2}\varphi^{(N)}
(\mathbf{x},t^{\prime\prime})]}{\delta\varepsilon(\mathbf{r}_{m}^{\prime},t^{\prime})}\right\}
\nonumber \\
+\frac{i}{2}\sum_{n=1}^{N}\xi_{n}\varphi^{(N)}(\mathbf{x},t^{\prime})\delta
(\mathbf{r}_{n}-\mathbf{r}_{m}^{\prime}) ,\label{FN_integral_2}
\end{gather}
where we have used the causality principle $\delta
\varphi^{(N)}(\mathbf{x},t^{\prime\prime})/\delta \varepsilon
(\mathbf{r}^{\prime},t^{\prime})=0$ for
$t^{\prime\prime}<t^{\prime}$,  (the factor $1/2$ appears due to
$\int_{0}^{\infty}\delta (t)dt$). In the limit
$t^{\prime}\rightarrow t$ we get
\begin{equation}
\label{var_der1}
 \frac{\delta \varphi^{(N)}(\mathbf{x},t)}{\delta
\varepsilon(\mathbf{r}_{m}^{\prime},t)}=\frac{i}{2}\sum_{n=1}^{N}\xi_{n}\varphi^{(N)}(\mathbf{x},t)\delta
(\mathbf{r}_{n}-\mathbf{r}_{m}^{\prime})
\end{equation}
Substituting Eq. (\ref{var_der1}) into Eq. (\ref{Furutsu1}), we
obtain
\begin{equation}
\langle\varepsilon(x_{m},t)\varphi^{(N)}\rangle=\frac{i}{2}\sum_{n=1}^{N}\xi_{n}
A(\mathbf{r}_{m}-\mathbf{r}_{n})\langle \varphi^{(N)}\rangle.
\end{equation}
Finally substituting this into Eq. (\ref{FN_average}) we have
\begin{gather}
i\frac{\partial\langle\varphi^{(N)}\rangle}{\partial
t}+\sum_{m=1}^{N}\xi_{m}\Delta_{m}\langle\varphi^{(N)}\rangle
+\sum_{m=1}^{N}\xi_{m}\langle|\varphi(\mathbf{r}_{m},t)|^{2}\varphi^{(N)}\rangle
\nonumber \\
+\frac{i}{2}\sum_{m=1}^{N}\sum_{n=1}^{N}\xi_{m}\xi_{n}A(\mathbf{r}_{m}-\mathbf{r}_{n})
\langle\varphi^{(N)}\rangle=0 . \label{FN_average1}
\end{gather}
Equation (\ref{FN_average1}) is not a closed equation for
$\langle\varphi^{(N)}\rangle$ due to the presence of
nonlinearities. An attempt to effect a closure of the third term
on the left-hand side of Eq. (\ref{FN_average1}) will result in
the infinite hierarchy of coupled moment equations mentioned in
the introduction.

In many physical applications, the correlation function
$A(\bm{\rho})$ is isotropic with the characteristic scale
$\rho_{c}$ and of power-law type for $|\bm{\rho}|<\rho_{c}$, so
that for the structure function $A(0)-A(|\bm{\rho}|)$ one can
write as
\begin{equation}
A(0)-A(|\bm{\rho}|)=A(0)(\rho/\rho_{c})^{\nu},
\end{equation}
where $\nu\leq 2$ (see, e. g. discussion in
Ref.~\cite{Jokipii1975}). The case $\nu=5/3$, in particular,
corresponds to the well-known Kolmogorov spectrum $\sim
\kappa^{-11/3}$, where $\kappa$ is the wave number. In what
follows, as in Ref.~\cite{Besieris_main}, we restrict ourselves to
the case $\nu=2$ which corresponds to a simplified or quadratic
Kolmogorov spectrum. Such a quadratic approximation of the
autocorrelation function occurs in many physically interesting
situations \cite{Ishimaru_book,Besieris1979,Bezak1998}. In
particular, the quadratic spectrum appears in the turbulence (e.
g. in turbulent atmosphere and ionosphere) on scales smaller than
the Kolmogorov length scale \cite{Ishimaru_book}. In addition,
strictly speaking, the use of such a quadratic form is justified
only if the functionals of $A(0)-A(|\bm{\rho}|)$  is not sensitive
at too large distances, which takes into account the exponential
localization of the structures considered below. Then we have
\begin{equation}
\label{spectr}
A(\mathbf{r}-\mathbf{r}^{\prime})=A(0)\left[1-\frac{(\mathbf{r}-\mathbf{r}^{\prime})^{2}}{\rho_{c}^{2}}\right],
\end{equation}
and further we denote $B=3A(0)/(8\rho_{c}^{2})$ (the factor $3/8$
is introduced to simplify the appearance of  expressions in
Sec.~VI). Then inserting Eq. (\ref{spectr}) into Eq.
(\ref{FN_average1}) one can obtain
\begin{gather}
i\frac{\partial\langle\varphi^{(N)}\rangle}{\partial
t}+\sum_{m=1}^{N}\xi_{m}\Delta_{m}\langle\varphi^{(N)}\rangle
+\sum_{m=1}^{N}\xi_{m}\langle|\varphi(\mathbf{r}_{m},t)|^{2}\varphi^{(N)}\rangle
\nonumber \\
-\frac{3i}{8}B\sum_{m<n}^{N}\sum_{n=1}^{N}\xi_{m}\xi_{n}(\mathbf{r}_{m}-\mathbf{r}_{n})^{2}
\langle\varphi^{(N)}\rangle=0. \label{FN_average2}
\end{gather}
This equation, like Eq. (\ref{FN_average1}), is of course also not
a closed equation for $\langle\varphi^{(N)}\rangle$.

\section{Effective nonlinear stochastic equation}

Let us consider the auxiliary nonlinear stochastic equation for
the function $\varphi_{\mathrm{e}}$,
\begin{equation}
\label{main11} i\frac{\partial\varphi_{\mathrm{e}}}{\partial
t}+\Delta\varphi_{\mathrm{e}}+|\varphi_{\mathrm{e}}|^{2}\varphi_{\mathrm{e}}
+\frac{1}{2}\mathbf{r}\cdot\mathbf{f}(t)\varphi_{\mathrm{e}}=0,
\end{equation}
where $\mathbf{f}(t)$ is a zero-mean, stationary,
$\delta$-correlated in time, vector real Gaussian random process
characterized by the correlation matrix
\begin{equation}
\label{correlator2} \langle f_{i}(t)f_{j}(t^{\prime})
\rangle=\frac{3}{2}B\delta_{ij}\,\delta (t-t^{\prime}),
\end{equation}
and $B$ is defined in Eq. (\ref{spectr}). In the one-dimensional
case, Eq. (\ref{main11}) was previously proposed in
Ref.~\cite{Besieris_main} (see also Eq. (2.3) in
Ref.~\cite{Bass1988}, and Eq. (3.2.4) in
Ref.~\cite{Konotop_book1994}). We define the function
$\varphi_{\mathrm{e}}^{(N)}$ in the same way as in Eq. (\ref{F})
by replacing $\varphi$ with $\varphi_{\mathrm{e}}$,
\begin{equation}
\varphi_{\mathrm{e}}^{(N)}(\mathbf{x},t)=\prod_{m=1}^{N/2}\varphi_{\mathrm{e}}(\mathbf{r}_{2m-1},t)
\varphi_{\mathrm{e}}^{\ast}(\mathbf{r}_{2m},t).
\end{equation}
Then, by analogy with Eq. (\ref{FN}), we get
\begin{gather}
i\frac{\partial\varphi_{\mathrm{e}}^{(N)}}{\partial
t}+\sum_{m=1}^{N}\xi_{m}\Delta_{m}\varphi_{\mathrm{e}}^{(N)}
+\sum_{m=1}^{N}\xi_{m}|\varphi_{\mathrm{e}}(\mathbf{r}_{m},t)|^{2}\varphi_{\mathrm{e}}^{(N)}
\nonumber \\
+\frac{1}{2}\sum_{m=1}^{N}\xi_{m}\mathbf{r}_{m}\cdot\mathbf{f}(t)\varphi_{\mathrm{e}}^{(N)}=0.
\label{FN_aux}
\end{gather}
Averaging Eq. (\ref{FN_aux}) yields
\begin{gather}
i\frac{\partial\langle\varphi_{\mathrm{e}}^{(N)}\rangle}{\partial
t}+\sum_{m=1}^{N}\xi_{m}\Delta_{m}\langle\varphi_{\mathrm{e}}^{(N)}\rangle
+\sum_{m=1}^{N}\xi_{m}\langle|\varphi_{\mathrm{e}}(\mathbf{r}_{m},t)|^{2}\varphi_{\mathrm{e}}^{(N)}\rangle
\nonumber \\
+\frac{1}{2}\sum_{m=1}^{N}\xi_{m}\mathbf{r}_{m}\cdot\langle\mathbf{f}(t)\varphi_{\mathrm{e}}^{(N)}\rangle=0,
\label{FN_aux_average}
\end{gather}
The statistical averaging in the fourth term on the left-hand side
of Eq. (\ref{FN_aux_average}) can be handled by the
Donsker-Furutsu-Novikov formula (\ref{Furutsu}). Then, using Eq.
(\ref{correlator2}), we have
\begin{equation}
\left\langle \mathbf{f}(t)\varphi_{\mathrm{e}}^{(N)}
\right\rangle=\frac{3}{2}B\left\langle\frac{\delta
\varphi_{\mathrm{e}}^{(N)}}{\delta \mathbf{f}(t)}\right\rangle
\end{equation}
The functional derivative can be found from Eq. (\ref{FN_aux}),
\begin{equation}
\frac{\delta \varphi_{\mathrm{e}}^{(N)}(\mathbf{x},t)}{\delta
\mathbf{f}(t)}=\frac{i}{4}\sum_{n=1}^{N}\xi_{n}\mathbf{r}_{n}\varphi_{\mathrm{e}}^{(N)}(\mathbf{x},t),
\end{equation}
and then we have
\begin{equation}
\label{aver1} \left\langle \mathbf{f}(t)\varphi_{\mathrm{e}}^{(N)}
\right\rangle=\frac{3i}{8}B\sum_{n=1}^{N}\xi_{n}\mathbf{r}_{n}
\langle\varphi_{\mathrm{e}}^{(N)}(\mathbf{x},t)\rangle .
\end{equation}
Substituting Eq. (\ref{aver1}) into Eq. (\ref{FN_aux_average}) we
have
\begin{gather}
i\frac{\partial\langle\varphi_{\mathrm{e}}^{(N)}\rangle}{\partial
t}+\sum_{m=1}^{N}\xi_{m}\Delta_{m}\langle\varphi_{\mathrm{e}}^{(N)}\rangle
+\sum_{m=1}^{N}\xi_{m}\langle|\varphi_{\mathrm{e}}(\mathbf{r}_{m},t)|^{2}\varphi_{\mathrm{e}}^{(N)}\rangle
\nonumber \\
+\frac{3i}{8}B\sum_{m=1}^{N}\sum_{n=1}^{N}\xi_{m}\xi_{n}\mathbf{r}_{m}\cdot\mathbf{r}_{n}
\langle\varphi_{\mathrm{e}}^{(N)}\rangle=0.
\label{FN_aux_average1}
\end{gather}
Using identity (it can be proved, for example, by the method of
mathematical induction)
\begin{equation}
-\sum_{m<n}^{N}\sum_{n=1}^{N}\xi_{m}\xi_{n}(\mathbf{r}_{m}
-\mathbf{r}_{n})^{2}=\sum_{m=1}^{N}\sum_{n=1}^{N}\xi_{m}\xi_{n}\mathbf{r}_{m}\cdot\mathbf{r}_{n},
\end{equation}
and comparing Eq. (\ref{FN_average2}) and Eq.
(\ref{FN_aux_average1}) we can see that the equations for
$\langle\varphi^{(N)}\rangle$ and
$\langle\varphi_{\mathrm{e}}^{(N)}\rangle$ are identical, that is
$\langle\varphi^{(N)}\rangle=\langle\varphi_{\mathrm{e}}^{(N)}\rangle$.
We emphasize that the equivalence of equations
(\ref{FN_aux_average1}) and (\ref{FN_average2}) is valid only
under the assumption of the quadratic Kolmogorov spectrum of the
random process $\varepsilon(\mathbf{r},t)$. Equation
(\ref{FN_aux_average1}), being the same with Eq.
(\ref{FN_average2}), is not a closed equation for
$\langle\varphi_{\mathrm{e}}^{(N)}\rangle$, but further we will
show that this equation, in contrast to Eq. (\ref{FN_average2}),
admits a global statistical analysis. We use the following ansatz
consisting in the change of variables
\begin{equation}
\label{change} \mathbf{R}(t)=\mathbf{r}-\mathbf{u}(t)
\end{equation}
and introducing of a new function $\psi(\mathbf{R},t)$ by
\begin{equation}
\label{sub}
\varphi_{\mathrm{e}}(\mathbf{r},t)=\psi(\mathbf{R}(t),t)\exp\{
i[\mathbf{u}_{t}(t)\cdot\mathbf{R}(t)/2+\gamma(t)]\},
\end{equation}
where the subscript denotes differentiation in $t$. By direct
substitution Eq. (\ref{sub})  into Eq. (\ref{main11})   one can
see that $\psi$ satisfy the equation
\begin{equation}
\label{mai1} i\frac{\partial\psi}{\partial
t}+\Delta_{\mathbf{R}}\psi+|\psi|^{2}\psi=0,
\end{equation}
where $\Delta_{\mathbf{R}}=\partial^{2}/\partial
R_{1}^{2}+\partial^{2}/\partial R_{2}^{2}++\partial^{2}/\partial
R_{3}^{2}$ (and the corresponding expression for the 2D case),
provided that the functions $\mathbf{u}(t)$ and $\gamma(t)$
satisfy equations
\begin{equation}
\label{langev1} \mathbf{u}_{tt}=\mathbf{f}(t), \,\,\,
\mathbf{u}(0)=0, \,\,\, \mathbf{u}_{t}(0)=0,
\end{equation}
and
\begin{equation}
\label{langev2}
\gamma_{t}=\frac{1}{4}\mathbf{u}^{2}_{t}(t)+\frac{1}{2}\mathbf{u}(t)\cdot\mathbf{f}(t),\,\,\,\gamma(0)=0
\end{equation}
for $t\geqslant 0$. The choice of zero initial data provides a
one-to-one mapping of the dependent variable $\psi$ onto
$\varphi_{e}$. Thus, the ansatz (\ref{sub})  has transformed the
stochastic nonlinear equations (\ref{main11})  into an explicitly
deterministic nonlinear equations for $\psi(\mathbf{R},t)$. It
should be noted, however, that $\psi$ is random function by virtue
of its dependence on $\mathbf{u}(t)$ through $\mathbf{R}(t)$ in
Eq. (\ref{change}). The solution $\psi$ of Eq. (\ref{mai1}) is
used  in the expression (\ref{sub}) for $\varphi_{e}$. This
function, in turn, is a functional of $\mathbf{u}$,
$\mathbf{u}_{t}$ and $\gamma$ because of its dependence on $\psi$
and the presence of the exponential factor in Eq. (\ref{sub}). The
computation of statistical moments of $\varphi_{e}$ require
knowledge of the joint probability density functional of the
random processes $\mathbf{u}$, $\mathbf{u}_{t}$ and $\gamma$.
Here, however, we are only interested in even $N$th order moments
$\langle\varphi_{\mathrm{e}}^{(N)}\rangle$ or, equivalently,
moments of Eq. (\ref{F}), and then $\varphi_{\mathrm{e}}$ is
independent of $\gamma$. Thus, our statistical analysis is limited
to finding only the joint density functional of $\mathbf{u}$ and
$\mathbf{u}_{t}$.

In the conclusion of this section we would like to make a remark
regarding Eq. (\ref{main11}) and its 1D version. It should be
noted that the replacements Eqs. (\ref{change}) and (\ref{sub})
reducing the non-autonomous equation (\ref{main11}) with an
explicit dependence on $t$ to the autonomous equation (\ref{mai1})
is rather nontrivial. In the 1D case, such a replacement was
proposed by Besieris in Ref.~\cite{Besieris_main}. On the other
hand, in the 1D case, as is known, Eq. (\ref{mai1}) is completely
integrable, has $N$-soliton solutions and can be solved by the
inverse scattering transform (IST) method
\cite{Zakharov-book1984}. Indeed, by generalizing the spectral
Zakharov-Shabat problem for the NLS equation, it was shown that
the one-dimensional version of Eq. (\ref{main11}) can be solved
using the IST   \cite{Serkin2001,Serkin2002}

\section{Fokker-Planck equation}

Under the assumptions made about the vector process $\mathbf{f}$,
equation (\ref{langev1}) with the external random force
$\mathbf{f}$ determined by the autocorrelation matrix
(\ref{correlator2}) is the equation for one Brownian particle and
it can be written as a system
\begin{eqnarray}
\mathbf{u}_{t}=\mathbf{v}, \label{langev3}
\\
\mathbf{v}_{t}=\mathbf{f}(t), \label{langev4}
\end{eqnarray}
with the initial conditions $\mathbf{v}(0)=0$ and
$\mathbf{u}(0)=0$. Then, the Fokker-Planck equation
\cite{Kampen-book1992} for the joint probability density function
$p(\mathbf{u},\mathbf{v},t)$ has the form
\begin{equation}
\label{Fokker} \frac{\partial p}{\partial
t}+\mathbf{v}\frac{\partial p}{\partial
\mathbf{u}}=\frac{3B}{4}\frac{\partial^{2} p}{\partial
\mathbf{v}^{2} }
\end{equation}
with the initial condition
$p(\mathbf{u},\mathbf{v},0)=\delta(\mathbf{u})\delta(\mathbf{v})$.
In the Appendix, we derived a more general Fokker-Planck equation
for a joint probability density function
$p(\mathbf{u},\mathbf{v},\gamma,t)$ that also depends on $\gamma$.
Equation (\ref{Fokker}) follows from (\ref{Fokker-general}) when
the dependence on $\gamma$ is neglected. It can readily be solved
and the solution has the form of the six-dimensional gaussian
distribution with the moments
\begin{equation}
\langle u_{i}u_{j}\rangle =\delta_{ij}\sigma_{11}^{2}, \,\,
\langle u_{i}v_{j}\rangle =\delta_{ij}\sigma_{12}^{2}, \,\,
\langle v_{i}v_{j}\rangle =\delta_{ij}\sigma_{22}^{2},
\end{equation}
where $i,j=x,y,z$, and
\begin{equation}
\sigma_{11}^{2}=\frac{1}{2}Bt^{3}, \,\,\,
\sigma_{12}^{2}=\frac{3}{4}Bt^{2}, \,\,\,
\sigma_{22}^{2}=\frac{3}{2}Bt.
\end{equation}
Separate components of $\mathbf{u}$ and $\mathbf{v}$ are
statistically independent due to Eqs. (\ref{correlator2}),
(\ref{langev3}) and (\ref{langev4}), so that the six-dimensional
gaussian distribution is the product of the two-dimensional
distributions for the each component,
\begin{equation}
p(\mathbf{u},\mathbf{v},t)=\prod_{i=1}^{D}p(u_{i},v_{i},t).
\end{equation}
For $p(u_{x},v_{x},t)$, for example, one obtains
\cite{Kampen-book1992}
\begin{gather}
p(u_{x},v_{x},t)=\frac{1}{2\pi\sqrt{\det\Sigma}} \nonumber \\
\times\exp\left[-\frac{1}{2\det\Sigma}
\left(\sigma_{22}^{2}u_{x}^{2}+\sigma_{11}^{2}v_{x}^{2}-2\sigma_{12}^{2}u_{x}v_{x}\right)\right],
\end{gather}
where $\Sigma$ is the $2\times 2$ covariance matrix with the
elements $\sigma_{mn}^{2}$, $m,n=1,2$ and
$\det\Sigma=3B^{2}t^{4}/16$. Then the joint probability density
function of $\mathbf{u}$ and $\mathbf{v}$ is
\begin{gather}
p(\mathbf{u},\mathbf{v},t)=\frac{1}{(2\pi)^{D}(\det\Sigma)^{D/2}} \nonumber \\
\times\exp\left[-\frac{1}{2\det\Sigma}
\left(\sigma_{22}^{2}\mathbf{u}^{2}+\sigma_{11}^{2}\mathbf{v}^{2}
-2\sigma_{12}^{2}\mathbf{u}\cdot\mathbf{v}\right)\right].
\label{puv}
\end{gather}
Note that in the limit $t\rightarrow\infty$ the joint probability
density function $p(\mathbf{u},\mathbf{v},t)$ tends to zero as
$\sim 1/t^{2D}$. From Eq. (\ref{puv}) it follows that
\begin{equation}
\label{p-u} p(\mathbf{u},t)=\int
p(\mathbf{u},\mathbf{v},t)\,d^{D}\mathbf{v}=\frac{1}{(2\pi)^{D/2}\sigma_{11}^{D}}
\exp\left(-\frac{\mathbf{u}^{2}}{2\sigma_{11}^{2}}\right),
\end{equation}
\begin{equation}
\label{p-v} p(\mathbf{v},t)=\int
p(\mathbf{u},\mathbf{v},t)\,d^{D}\mathbf{u}=\frac{1}{(2\pi)^{D/2}\sigma_{22}^{D}}
\exp\left(-\frac{\mathbf{v}^{2}}{2\sigma_{22}^{2}}\right).
\end{equation}
In the limit $t\rightarrow\infty$, the probability density
functions $p(\mathbf{u},t)$ and $p(\mathbf{v},t)$ behave as $\sim
1/t^{3D/2}$ and $\sim 1/t^{D/2}$, respectively. Given the known
joint probability density function $p(\mathbf{u},\mathbf{v},t)$,
we can write
\begin{equation}
\langle\varphi^{(N)}\rangle=\langle\varphi^{(N)}_{\mathrm{e}}\rangle=\int
\varphi^{(N)}_{\mathrm{e}}
p(\mathbf{u},\mathbf{v},t)d^{D}\mathbf{u}\,d^{D}\mathbf{v},
\end{equation}
where $\varphi^{(N)}_{\mathrm{e}}$ is constructed from solutions
of Eq. (\ref{mai1}) using Eqs. (\ref{change}) and (\ref{sub}).

\section{Deterministic stationary solutions}

In this section we will consider stationary solutions of Eq.
(\ref{mai1}) in the form of multidimensional coherent structures,
namely the fundamental soliton and the vortex soliton. Since
corresponding exact analytical solutions do not exist (and
numerical solutions are not adequate for further analysis), we use
the variational method. Note that these solutions are purely
formal in the sense that here we do not study the problem of the
stability of such solutions. Equation (\ref{mai1}) can be written
in Hamiltonian form
\begin{equation}
i\frac{\partial \psi}{\partial t}=\frac{\delta
H}{\delta\psi^{\ast}},
\end{equation}
where the Hamiltonian is
\begin{equation}
\label{H} H=\int
\left(|\nabla\psi|^{2}-\frac{|\psi|^{4}}{2}\right)d^{D}\mathbf{R}.
\end{equation}
Other integrals of motion are the norm (also called energy, number
of particles, or in optics wave power)
\begin{equation}
\label{N} N=\int |\psi|^{2}d^{D}\mathbf{R},
\end{equation}
the momentum
\begin{equation}
\label{P}
\mathbf{P}=\frac{i}{2}\int(\psi\nabla\psi^{\ast}-\psi^{\ast}\nabla\psi)d^{D}\mathbf{R}
\end{equation}
and the angular momentum
\begin{equation}
\label{M} \mathbf{M}=i\int \mathbf{R}\times
(\psi^{\ast}\nabla\psi-\psi\nabla\psi^{\ast})d^{D}\mathbf{R}.
\end{equation}
An extensive literature is devoted to the NLS equation
(\ref{mai1}) with the dimension of space $D$ (see, for example,
\cite{Sulem1999} and numerous references therein). It is well
known that the NLS equation with $D = 1$ is completely integrable
and admits $N$-soliton solutions \cite{Zakharov-book1984}, while
for $D\geqslant 2$ the initial field distribution undergoes
collapse (or self-focusing if $t$ has the meaning of a spatial
variable along the beam)  or spreads out
\cite{Zakharov_UFN2012,Fibich_book,Berge1998}. The collapse is
critical if $D=2$, which means that it sets in when the norm of
the underlying wave field exceeds a certain finite critical value,
and supercritical in 3D, where an initial state with an
arbitrarily small norm may blow up. Under this, the initial
perturbation in any case never reaches a stationary state (in
other words, stationary solutions are unstable). We are still
looking for formal stationary solutions of Eq. (\ref{mai1}) of the
form
\begin{equation}
\label{stat-sol} \psi (\mathbf{R},t)=\phi
(\mathbf{R})\mathrm{e}^{i\lambda t}.
\end{equation}
These solutions satisfy the equation
\begin{equation}
\label{stat-eq}
-\lambda\phi+\Delta_{\mathbf{R}}\phi+|\phi|^{2}\phi=0,
\end{equation}
and resolve the variational problem $\delta S = 0$ for the
functional
\begin{equation}
\label{S} S=H+\lambda N.
\end{equation}
Note that in this paper we restrict ourselves to motionless and
non-rotating solutions. Solutions moving with the constant
velocity $\mathbf{v}$ correspond to the additional term
$\mathbf{v}\cdot\mathbf{P}$ in the functional $S$
\cite{Berge2002}. Due to the Galilean invariance of Eq.
(\ref{mai1}), moving solutions are easily obtained from motionless
solutions by the appropriate Galilean transformation. For
solutions rotating with the constant angular velocity
$\bm{\omega}$, i.e., which are stationary in the reference frame
rotating with constant angular velocity $\bm{\omega}$, the
functional $S$ would also contain term
$\bm{\omega}\cdot\mathbf{M}$. The procedure for finding such
solutions (both by variational and direct numerical methods) is
described in \cite{Lashkin2012} and these solutions are vortex
clusters (azimuthon solitons). In this paper, we do not consider
such rotating stationary structures, although the developed
formalism is fully applicable to them as well.

The solution of Eq. (\ref{stat-eq}) in the form of a fundamental
soliton (ground state) corresponds to the radially symmetric case
with boundary conditions $\partial\phi/\partial R=0$ at zero $R=0$
and $\phi\rightarrow 0$ at infinity $R\rightarrow\infty$. The
vortex soliton corresponds to ansatz $\phi=\phi (R)\exp
(im\theta)$, where $\theta$ is the polar angle and $m$ is an
integer (topological charge), with boundary conditions $\phi=0$ at
zero $R=0$ and $\phi\rightarrow 0$ at infinity
$R\rightarrow\infty$. The important integral of motion associated
with this type of solitary wave is the angular momentum. One can
show that for the vortex soliton $M_{z}=mN$. Solutions of Eq.
(\ref{stat-eq}) in the form of fundamental and vortex solitons can
be found numerically in both 2D and 3D cases
\cite{Malomed_vortex2019}, and the corresponding solutions turn
out to be unstable.

For an analytical treatment, following Ref.~\cite{Berge2002}, we
consider the trial function of the form
$\phi(\mathbf{R})=A\phi_{a}(\mathbf{R}/a)$, where $A$ and $a$ are
unknown parameters to be determined by the variational procedure
from the system of equations $\partial S/\partial A=0$ and
$\partial S/\partial a=0$. For not too large values of $\lambda$,
or equivalently $N$, the variational analysis is in good agreement
with the solutions obtained by exact numerical simulation. We
denote $\bm{\xi}=\mathbf{R}/a$ and for an arbitrary form of the
$\phi_{a}(\bm{\xi})$, provided $\phi_{a}$ vanishes at infinity
fast enough so that the corresponding integrals are convergent,
one can get
\begin{equation}
\label{param} A^{2}=\frac{2\lambda\delta}{\beta (4-D)}, \quad
a^{2}=\frac{\alpha (4-D)}{D\lambda\delta},
\end{equation}
where
\begin{gather}
\alpha=\int |\nabla\varphi_{a}|^{2}d^{D}\bm{\xi}, \quad \beta=\int
\frac{|\varphi_{a}|^{4}}{2}d^{D}\bm{\xi},
\label{alpha} \\
\label{delta} \delta=\int |\varphi_{a}|^{2}d^{D}\bm{\xi}.
\end{gather}
We choose the test functions in the form:
\begin{description}
\item[\normalfont{(i)}] The ground state (fundamental soliton)
\begin{equation}
\label{soliton-trial} \phi_{a}(\bm{\xi})=\exp (-\xi^{2}/2).
\end{equation}
\item[\normalfont{(ii)}]
The vortex soliton
\begin{equation}
\label{vortex-trial} \phi_{a}(\bm{\xi})=\xi^{|m|}\exp
(-\xi^{2}/2)\mathrm{e}^{im\theta},
\end{equation}
\end{description}
and further we restrict ourselves to the case $m=1$. Substituting
Eqs. (\ref{soliton-trial}) and (\ref{vortex-trial}) into Eqs.
(\ref{alpha}) and (\ref{delta}), and calculating the corresponding
integrals one can obtain, following Ref.~\cite{Berge2002}, for the
fundamental soliton in the 2D case $\alpha=\delta=\pi$, $\beta=\pi
/4$, and in the 3D case $\alpha=3\pi^{3/2}/2$,
$\beta=\sqrt{2}\pi^{3/2}/8$, $\delta=\pi^{3/2}$. For the vortex
soliton we have $\alpha=2\pi$,$\beta=\pi/8$, $\delta=\pi$ in the
2D case, and $\alpha=5\pi^{3/2}/2$,$\beta=\sqrt{2}\pi^{3/2}/16$,
$\delta=\pi^{3/2}$ in the 3D case. Then, for the parameters $A$
(the amplitude) and $a$ (characteristic size of the structures),
from Eq. (\ref{param}) we can obtain for the fundamental soliton,
\begin{equation}
\label{A2-sol}
A^{2}=\left\{ \begin{array}{ll} 4\lambda & \text{for 2D} , \\
8\sqrt{2}\lambda & \text{for 3D}
\end{array}
\right. , \quad
a^{2}=\left\{ \begin{array}{ll} 1/\lambda & \text{for 2D} , \\
1/(2\lambda) & \text{for 3D}
\end{array}
\right. ,
\end{equation}
and for the vortex soliton,
\begin{equation}
\label{A2-vor}
A^{2}=\left\{ \begin{array}{ll} 8\lambda & \text{for 2D} , \\
16\sqrt{2}\lambda & \text{for 3D}
\end{array}
\right. , \quad
a^{2}=\left\{ \begin{array}{ll} 2/\lambda & \text{for 2D} , \\
5/(6\lambda) & \text{for 3D}
\end{array}
\right. .
\end{equation}
Since $N=A^{2}a^{D}\delta$, it follows from Eq. (\ref{A2-sol})
that for the 3D fundamental soliton $\partial N/\partial\lambda
<0$, that is, according to the Vakhitov-Kolokolov criterion, the
soliton is unstable. For a two-dimensional soliton, we have
$\partial N/\partial\lambda =0$, which corresponds to the marginal
instability (critical collapse).

\section{Averaged observables for fundamental and vortex solitons}

Due to the parametric dependence of $\mathbf{R}$ on time in Eq.
(\ref{change}), solutions (\ref{stat-sol}), (\ref{soliton-trial})
and (\ref{vortex-trial}) of Eq. (\ref{mai1}) presented in the
previous section with replacement Eq. (\ref{sub}) are not
stationary at all and, moreover, are random functions. In this
case they have the form
\begin{equation}
\label{soliton}  \varphi_{\mathrm{e}}(\mathbf{r},t)
=A\exp\left[-\frac{(\mathbf{r}-\mathbf{u})^{2}}{2a^{2}}+i\frac{\mathbf{v}\cdot
(\mathbf{r}-\mathbf{u})}{2a^{2}}+i\gamma+i\lambda t\right]
\end{equation}
for the fundamental soliton, and
\begin{gather}
\varphi_{\mathrm{e}}(\mathbf{r},t)=A\frac{[(x-u_{x})+i(y-u_{y})]}{a}
\nonumber \\
\times\exp\left[-\frac{(\mathbf{r}-\mathbf{u})^{2}}{2a^{2}}+i\frac{\mathbf{v}\cdot
(\mathbf{r}-\mathbf{u})}{2a^{2}}+i\gamma+i\lambda t\right]
\label{vortex}
\end{gather}
for the vortex soliton, respectively. Our first task is to compute
the mean intensities $\langle |\varphi(\mathbf{r},t)|^{2}
\rangle=\langle |\varphi_{\mathrm{e}}(\mathbf{r},t)|^{2} \rangle$
of the fundamental and vortex solitons. Next we denote
$I(\mathbf{r},t)=|\varphi(\mathbf{r},t)|^{2}$. In this case the
function $I(\mathbf{r},t)$ is a function of only of the random
process $\mathbf{u}(t)$. Its ensemble average, therefore, given by
\begin{equation}
\label{intensity-both} \langle I(\mathbf{r},t)\rangle =\int
p(\mathbf{u},t)I(\mathbf{r},t)d^{D}\mathbf{u},
\end{equation}
where $p(\mathbf{u},t)$ is determined by Eq. (\ref{p-u}) and
$I(\mathbf{r},t)$ is determined from Eqs. (\ref{soliton}) and
(\ref{vortex}) for the fundamental and vortex soliton
respectively.
\begin{figure}
\includegraphics[width=3.2in]{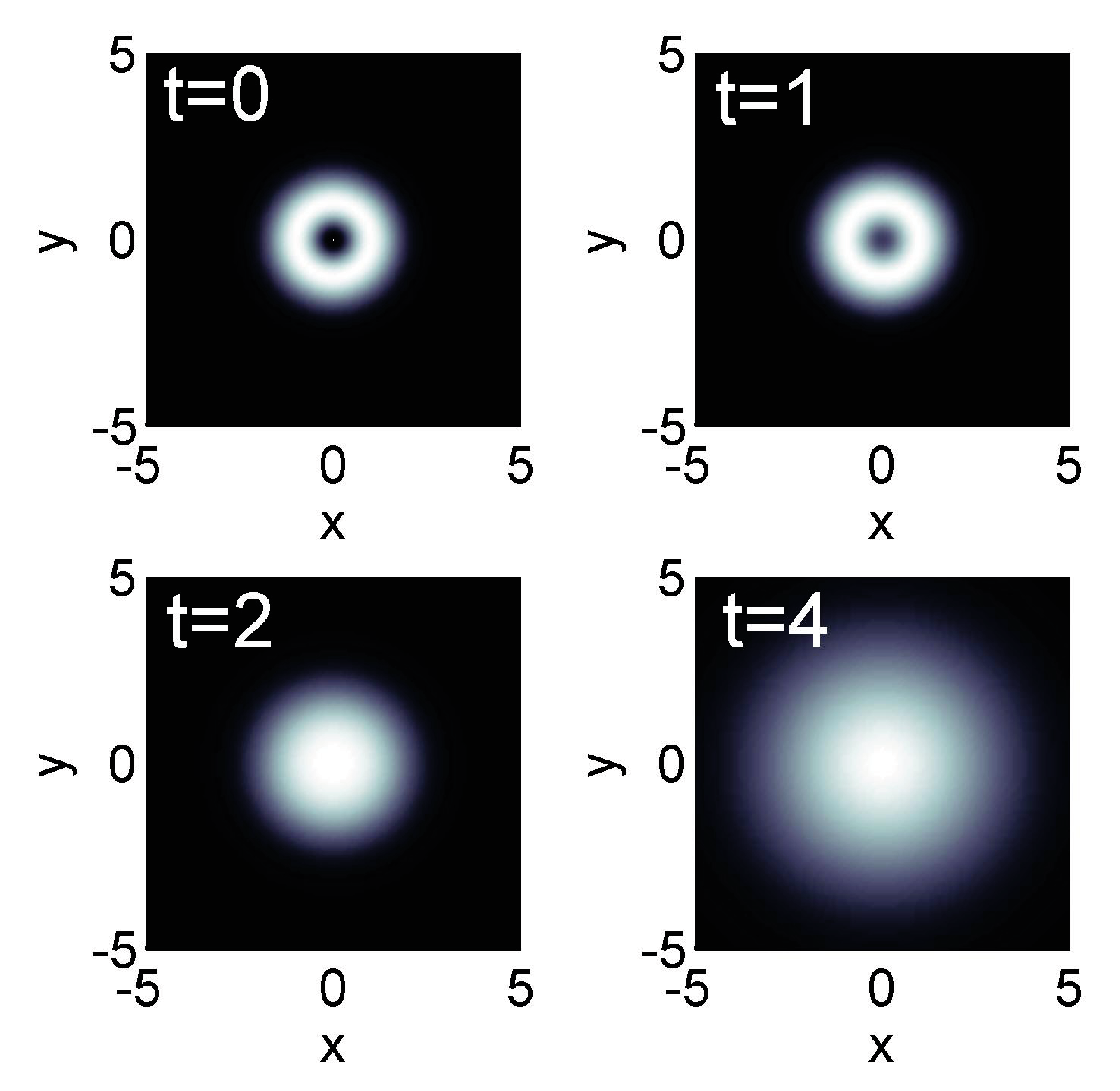}
\caption{\label{fig1} The mean 2D vortex soliton intensity
$\langle I(\mathbf{r},t)\rangle$ in the presence of random
fluctuations according to the analytical prediction
Eq.~(\ref{intensity-vortex})  for the parameters $\lambda=1$ and
$B=0.05$.}
\end{figure}
Calculating the integrals in Eq.~(\ref{intensity-both}), we obtain
for the fundamental soliton,
\begin{equation}
\label{intensity-soliton} \langle
I(\mathbf{r},t)\rangle=\frac{A^{2}a^{D}}{(a^{2}+Bt^{3})^{D/2}}
\exp \left[-\frac{r^{2}}{(a^{2}+Bt^{3})}\right],
\end{equation}
where $A$ and $a$ are determined by Eq. (\ref{A2-sol}), and for
the vortex soliton
\begin{gather}
\langle I(\mathbf{r},t)\rangle=\frac{A^{2}a^{D}[r_{\perp}^{2}a^{2}
+Bt^{3}(a^{2}+Bt^{3})]}{(a^{2}+Bt^{3})^{D/2+2}} \nonumber \\
\times\exp \left[-\frac{r^{2}}{(a^{2}+Bt^{3})}\right],
\label{intensity-vortex}
\end{gather}
where $A$ and $a$ are determined by Eq. (\ref{A2-vor}). Note that
in the presence of randomness, as can be seen from
(\ref{intensity-vortex}), the zero at the center of the vortex
soliton disappears. In both cases the mean intensity spreads out
monotonically due to irreversible effects of randomness, and no
collapse occurs. The mean 2D vortex soliton intensity $\langle
I(\mathbf{r},t)\rangle$ with the parameters $\lambda=1$, $B$=0.05
for different times is shown in Fig.~1. As can be seen, the
conservative equation (\ref{main}) conserves  the norm $\int
|\varphi|^{2}d^{D}\mathbf{r}$ for any random realization of
$\varepsilon (\mathbf{r},t)$ and then the averaged norm
\begin{equation}
\langle N\rangle=\int \langle I(\mathbf{r},t)\rangle
d^{D}\mathbf{r}
\end{equation}
is also conserved. For analytical expressions obtained by the
variational method in the previous section, we have $\langle
N\rangle=A^{2}a^{D}\pi^{D/2}$. Next, we calculate the mean
momentum density
$\mathbf{p}=i(\psi\nabla\psi^{\ast}-\psi^{\ast}\nabla\psi)/2$ and
the mean angular momentum density $\mathbf{m}=[\mathbf{r}\times
\mathbf{p}]$. These quantities are nonzero only for a vortex
soliton, since here we consider motionless  structures. Then we
have
\begin{equation}
\langle \mathbf{p}(\mathbf{r},t)\rangle=\int
p(\mathbf{u},t)\mathbf{p}(\mathbf{r},t)d^{D}\mathbf{r},
\end{equation}
and calculating integrals we find for the mean  momentum density,
\begin{gather}
\langle
\mathbf{p}(\mathbf{r},t)\rangle=\frac{A^{2}a^{D}[\hat{\mathbf{z}}\times\mathbf{r}]}
{(a^{2}+Bt^{3})^{D/2+1}} \exp
\left[-\frac{r^{2}}{(a^{2}+Bt^{3})}\right], \label{flux-vortex}
\end{gather}
and for the mean angular momentum density we have $\langle
\mathbf{m}(\mathbf{r},t)\rangle=[\mathbf{r}\times \langle
\mathbf{p}(\mathbf{r},t)\rangle]$.

The mean centroid of the structure, defined by
\begin{equation}
\label{mean-centroid} \mathbf{\bar{r}}(t)=\frac{1}{\langle
N\rangle}\int \mathbf{r}\langle I(\mathbf{r},t)\rangle
d^{D}\mathbf{r},
\end{equation}
as follows from Eqs. (\ref{intensity-soliton}) and
(\ref{intensity-vortex}), is zero for all $t$. A measure of the
spreading of the structure can be found by using the definition
\begin{equation}
\label{r2}  \mathbf{r}^{2}_{s}(t) =\frac{1}{\langle N\rangle}\int
[\mathbf{r}-\mathbf{\bar{r}}(t)]^{2}\langle I(\mathbf{r},t)\rangle
d^{D}\mathbf{r}.
\end{equation}
Calculating integrals and taking into account that
$\mathbf{\bar{r}}(t)=0$ we find that the standard deviation
evolves in $t$ as
\begin{equation}
\label{r2-soliton} [\mathbf{r}^{2}_{s}(t)]^{1/2}=
\left[\frac{D}{2}(a^{2}+Bt^{3})\right]^{1/2}
\end{equation}
for the fundamental soliton, and
\begin{equation}
\label{r2-vortex}
[\mathbf{r}^{2}_{s}(t)]^{1/2}=\left[\frac{(D+2)a^{2}+DBt^{3}}{2}\right]^{1/2}
\end{equation}
for the vortex soliton. Dependence of $(\mathbf{r}_{s}^{2})^{1/2}$
on $t$ (recall that the evolutionary variable $t$ for 2D beams
plays the role of a spatial coordinate along the beam propagation)
of the form $\sim (\mathrm{const}+t^{3})^{1/2}$ is in agreement
with the result of Ref.~\cite{Gaididei1998} obtained with the aid
of the Donsker-Furutsu-Novikov formalism and virial theorem for
the 2D NLS equation. It is interesting to note that the same
$t$-dependence of the 2D beam position standard deviation was
found experimentally with high accuracy in an stochastic nonlocal
focusing Kerr medium such as liquid crystals
\cite{Louis-Experiment2016}.

Consider then the spatial mutual coherence function $\Gamma
(\mathbf{r}_{1},\mathbf{r}_{2},t)=\langle\varphi^{(2)}(\mathbf{r}_{1},\mathbf{r}_{2},t)\rangle$
that characterizes the angular spread and coherence properties of
the wave field and is defined as
\begin{equation}
\label{mutual} \Gamma (\mathbf{r}_{1},\mathbf{r}_{2},t)=\int
p(\mathbf{u},\mathbf{v},t)\varphi (\mathbf{r}_{1},t)\varphi^{\ast}
(\mathbf{r}_{2},t)d^{D}\mathbf{u}\,d^{D}\mathbf{v}.
\end{equation}
This function is important and is especially often used when
studying wave beams in optics in the presence of a random addition
to the refractive index. One can see that the function $\varphi
(\mathbf{r}_{1},t)\varphi^{\ast} (\mathbf{r}_{2},t)$ depends on
both $\mathbf{u}$ and $\mathbf{v}$. For the fundamental soliton we
have
\begin{gather}
\varphi (\mathbf{r}_{1},t)\varphi^{\ast}
(\mathbf{r}_{2},t)=A^{2}\exp\left[-\frac{(\mathbf{r}_{1}-\mathbf{u})^{2}
+(\mathbf{r}_{2}-\mathbf{u})^{2}}{2a^{2}} \right.
\nonumber \\
\left.
+i\frac{\mathbf{v}\cdot(\mathbf{r}_{1}-\mathbf{r}_{2})}{2a^{2}}\right],
\label{mutual-soliton}
\end{gather}
and for the vortex soliton
\begin{gather}
\varphi (\mathbf{r}_{1},t)\varphi^{\ast}
(\mathbf{r}_{2},t)=\frac{A^{2}}{a^{2}}[x_{1}-u_{x}+i(y_{1}-u_{y})][x_{2}-u_{x}
\nonumber
\\
+i(y_{2}-u_{y})]
\exp\left[-\frac{(\mathbf{r}_{1}-\mathbf{u})^{2}+(\mathbf{r}_{2}-\mathbf{u})^{2}}{2a^{2}}
\right.
 \nonumber
\\
\left.
+i\frac{\mathbf{v}\cdot(\mathbf{r}_{1}-\mathbf{r}_{2})}{2a^{2}}\right].
\label{mutual-vortex}
\end{gather}
To avoid a rather cumbersome expression, we present the
corresponding expression for the spatial mutual coherence function
only for the fundamental soliton. Using Eqs. (\ref{puv}) and
(\ref{mutual}) we get
\begin{widetext}
\begin{equation}
\label{Mutual}
\Gamma(\mathbf{r}_{1},\mathbf{r}_{2},t)=\frac{A^{2}a^{D}}{(a^{2}+Bt^{3})^{D/2}}
\exp\left\{-\frac{Bt(3a^{2}+4a^{2}t^{2}+3Bt^{3}/4)(\mathbf{r}_{1}-\mathbf{r}_{2})^{2}+8a^{4}
(\mathbf{r}_{1}^{2}+\mathbf{r}_{2}^{2})-6iBa^{2}t^{2}(\mathbf{r}_{1}^{2}
-\mathbf{r}_{2}^{2})}{16a^{4}(a^{2}+Bt^{3}) }\right\}.
\end{equation}
\end{widetext}
For $\mathbf{r}_{1}=\mathbf{r}_{2}$ we recover the result Eq.
(\ref{intensity-soliton}). From the known
$\Gamma(\mathbf{r}_{1},\mathbf{r}_{2},t)$, one can find the mean
Wigner distribution density function which is defined as follows,
\cite{Wigner1932}
\begin{equation}
\label{Wigner1}
W(\mathbf{p},\mathbf{r},t)=\frac{1}{(2\pi)^{D}}\int
\mathrm{e}^{i\mathbf{p}\cdot\mathbf{y}}\Gamma
\left(\mathbf{r}+\frac{\mathbf{y}}{2},\mathbf{r}-\frac{\mathbf{y}}{2},t\right)d^{D}\mathbf{y},
\end{equation}
and has numerous applications not only in statistical mechanics,
but also in quantum and classical optics, signal processing theory
etc. . This quantity, as is well known, is real, but not
necessarily positive everywhere. Provided that
$W(\mathbf{p},\mathbf{r},t)$ is normalized to unity, this function
is different from zero in a region of which the volume in phase
space is at least equal to $(2\pi)^{D}$ and
$W(\mathbf{p},\mathbf{r},t)$ can never be sharply localized in
$\mathbf{r}$ and $\mathbf{p}$. Inserting Eq. (\ref{Mutual}) into
Eq. (\ref{Wigner1}) and calculating integrals we have
\begin{widetext}
\begin{gather}
W(\mathbf{p},\mathbf{r},t)=\frac{2^{D}A^{2}a^{3D}}{\pi^{D+1/2}
[Bt(3a^{2}+4a^{2}t^{2}+3Bt^{3}/4)+4a^{4}]^{D/2}}
\nonumber \\
\times\exp\left\{-\frac{\mathbf{r}^{2}}{a^{2}+Bt^{3}}
-\frac{4a^{4}(a^{2}+Bt^{3})}{[Bt(3a^{2}+4a^{2}t^{2}+3Bt^{3}/4)+4a^{4}]}
\left[\frac{3Bt^{2}}{4a^{2}(a^{2}+Bt^{3})}\mathbf{r}+\mathbf{p}\right]^{2}\right\}.
\label{Wigner2}
\end{gather}
\end{widetext}

Let us consider the higher moment orders. We define the centroid
of the structure by
\begin{equation}
\mathbf{r}_{c}(t)=\frac{1}{\langle N \rangle}\int
\mathbf{r}I(\mathbf{r},t)d^{D}\mathbf{r}.
\end{equation}
As we have seen from Eq. (\ref{mean-centroid}), the mean centroid
$\mathbf{\bar{r}}(t)=\langle\mathbf{r}_{c}(t)\rangle$ have found
to be equal to zero for both fundamental and vortex soliton. It is
interesting, however, to consider also the variance of the random
function $\mathbf{r}_{c}(t)$,
\begin{gather}
\left\langle [\mathbf{r}_{c}(t)-\mathbf{\bar{r}}(t)]^{2}
\right\rangle=\int
p(\mathbf{u},t)[\mathbf{r}_{c}(t)-\mathbf{\bar{r}}(t)]^{2}
d\mathbf{u}
 \nonumber \\
 =\frac{1}{\langle N \rangle^{2}}\int\int \mathbf{r}_{1}\cdot \mathbf{r}_{2} p(\mathbf{u},t) I(\mathbf{r}_{1},t) I(\mathbf{r}_{2},t)
d^{D}\mathbf{u}\,d^{D}\mathbf{r}_{1}d^{D}\mathbf{r}_{2}.
\label{varHigh}
\end{gather}
Integration in Eq. (\ref{varHigh}) taking into account Eqs.
(\ref{p-u}), (\ref{intensity-soliton}) and
(\ref{intensity-vortex}) yields
\begin{equation}
\label{varian2} \left\langle
[\mathbf{r}_{c}(t)-\mathbf{\bar{r}}(t)]^{2} \right\rangle=CBt^{3},
\end{equation}
where $C=1$ and $C=3/2$ for the fundamental soliton in the 2D and
3D cases, respectively. For the vortex soliton, we have $C=2$ and
$C=27/8$, respectively.
\begin{figure}
\includegraphics[width=3.0in]{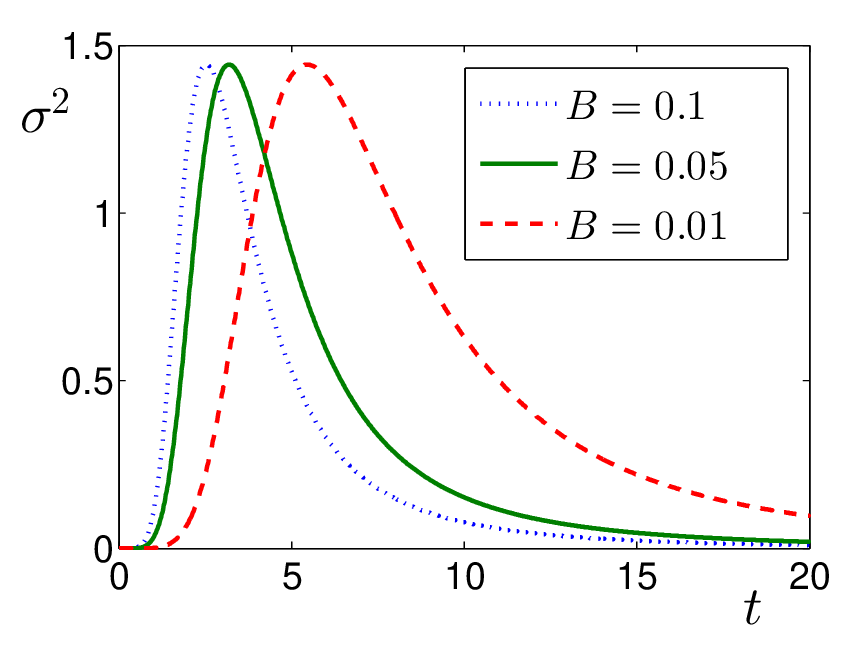}
\caption{\label{fig2} Time dependence of the variance of intensity
at the center $\sigma^{2}$ of the fundamental soliton for
$\lambda=1$ and different values of $B$ (noise intensity).}
\end{figure}
The quantity (\ref{varian2}) becomes unbounded as
$t\rightarrow\infty$. This indicates the stochastic instability of
these structures due to the presence of random fluctuations. Since
most experiments measure the intensity and not the field itself,
intensity fluctuations are of particular interest. An important
characteristic in this case is the variance of the intensity
\begin{equation}
\sigma^{2}_{I}(\mathbf{r},t)\equiv\left\langle
[I(\mathbf{r},t)-\langle
I(\mathbf{r},t)\rangle]^{2}\right\rangle=\langle
I^{2}(\mathbf{r},t)\rangle-\langle I(\mathbf{r},t)\rangle^{2}.
\end{equation}
The normalized variance $\sigma^{2}_{I}/\langle I \rangle^{2}$ is
commonly referred to as the scintillation (flicker) index
\cite{Tatarski_book1967,Ishimaru_book}. The calculations can be
carried out without difficulty and we have for the fundamental
soliton
\begin{gather}
\sigma^{2}_{I}(\mathbf{r},t)=A^{4}\left\{\frac{a^{D}\exp
\left[-2r^{2}/(a^{2}+2Bt^{3})\right]}{(a^{2}+2Bt^{3})^{D/2}}
\right.
\nonumber \\
\left.
 -\frac{a^{2D}\exp
\left[-2r^{2}/(a^{2}+Bt^{3})\right]}{(a^{2}+Bt^{3})^{D}}\right\},
\label{var-in-sol}
\end{gather}
and for the vortex soliton,
\begin{gather}
\sigma^{2}_{I}(\mathbf{r},t)=A^{4}\left\{a^{D+4}\exp
\left[-\frac{2r^{2}}{(a^{2}+2Bt^{3})}\right] \right.
\nonumber \\
\times\frac{[a^{2}r_{\perp}^{2}+2Bt^{3}(a^{2}+2Bt^{3})]^{2}
-2B^{2}t^{6}(a^{2}+2Bt^{3})^{2}}{(a^{2}+2Bt^{3})^{D/2+4}}
\nonumber \\
\left. -a^{2(D+2)}\exp
\left[-\frac{2r^{2}}{(a^{2}+Bt^{3})}\right]\frac{[a^{2}r_{\perp}^{2}
+Bt^{3}(a^{2}+Bt^{3})]^{2}}{(a^{2}+Bt^{3})^{D+4}}\right\}.
\label{var-in-vor}
\end{gather}
The variance of the intensity at the center of the fundamental and
vortex solitons can be easily found from Eqs. (\ref{var-in-sol})
and (\ref{var-in-vor}), and have the form
\begin{equation}
\sigma^{2}_{I}(0,t)=A^{4}\left\{\frac{a^{D}}{(a^{2}+2Bt^{3})^{D/2}}
 -\frac{a^{2D}}{(a^{2}+Bt^{3})^{D}}\right\}.
\end{equation}
and
\begin{equation}
\sigma^{2}_{I}(0,t)=A^{4}\left\{\frac{2a^{D+4}B^{2}t^{6}}{(a^{2}+2Bt^{3})^{D/2+2}}
 -\frac{a^{2(D+2)}B^{2}t^{6}}{(a^{2}+Bt^{3})^{D+2}}\right\}.
\end{equation}
respectively. The function $\sigma^{2}_{I}(0,t)$ in both cases is
a non-monotonic function of time and reaches a maximum at some
fixed moment of time, which increases with decreasing noise
intensity $B$. In the limit $t\rightarrow\infty$, the variance of
the intensity at the center tends to zero. Time dependence of the
variance of intensity at the center of the fundamental soliton for
$\lambda=1$ and different values of $B$ is shown in Fig.~2.

\section{Conclusion}

We have presented a new technique for calculating the even moments
of the $N$th order for the NLS equation with the spatial dimension
$D\geqslant 2$ with a multiplicative random term depending on the
time and space coordinates. The proposed formalism does not use
closure procedures for the nonlinear term, as well as the
smallness of the random term and the use of perturbation theory.
Along with the original Eq. (\ref{main}), we have considered an
auxiliary stochastic nonlinear equation. We have derived equations
(nonclosed) for the even moments of the $N$th order both for the
original and for the auxiliary (effective) equation. Assuming that
the autocorrelation function of the random field is quadratic, we
have shown that the even moments of the original equation and the
effective one coincide. By  a stochastic change of variables and a
fundamental ansatz, we have demonstrated how  the effective
stochastic nonlinear equation can be mapped into an explicit
deterministic equation for a new dependent variable and a system
of stochastic nonlinear ordinary differential equations. From this
system of Langevin-type equations, we have derived the
corresponding Fokker-Planck equation for the joint probability
density function and obtained its Gaussian-type solution. The even
moments of the field are then expressed in terms of the integrals
of the product of the deterministic fields and the corresponding
probability density function. As a specific example, we have used
analytical expressions for the fundamental and vortex solitons,
obtained using variational analysis. We have analytically
calculated a number of statistical characteristics describing the
dynamics of fundamental and vortex solitons in random medium, such
as the mean intensities, the variance of the intensity, the
centroid and spread of the structures, the spatial mutual
coherence function etc. In particular, we have shown that, under
the irreversible action of fluctuations, the solitons spread out,
i.e., no collapse (beam self-focusing) occurs.

It is important to note that for specific realizations (there are
an infinite number of them) of noise, the multidimensional
fundamental and vortex solitons are unstable and collapse. In this
regard, it is necessary to emphasize that the arrest of the
collapse of the considered coherent structures is understood only
in the sense of average values (more precisely, even moments)
when, for example, analytical expressions for the average width of
the structure clearly indicate spreading, and not collapse (beam
self-focusing).

Note that the developed method is also applicable to odd moments,
and of particular interest is the calculation of the first moment,
that is, the mean field $\langle\varphi\rangle$. This requires a
joint probability density function
$p(\mathbf{u},\mathbf{v},\gamma)$. Then, in particular, we have
for the mean field
\begin{equation}
\langle\varphi(\mathbf{r},t)\rangle=\int
\varphi_{\mathrm{e}}(\mathbf{r},t) p(\mathbf{u},\mathbf{v},\gamma,
t)d^{D}\mathbf{u}\,d^{D}\mathbf{v}d\gamma,
\end{equation}
where $\varphi_{\mathrm{e}}$ is determined by Eqs. (\ref{soliton})
or (\ref{vortex}). The equation for
$p(\mathbf{u},\mathbf{v},\gamma)$ can readily be obtained (see
Appendix). However, the analytical solution of the resulting
equation seems to encounter significant difficulties.

The formalism developed in this paper can also be applied to other
problems that can be formulated in terms of the model of the
stochastic NLS equation (\ref{main}). We emphasize, however, that
this approach  is applicable only for a very specific type
(quadratic spatial dependence) of spatial noise correlator.

\section*{Appendix: derivation of the extended Fokker-Planck equation}

The microscopic ("fine-grained") distribution function associated
to the system (\ref{langev2})-(\ref{langev4}), is
\begin{gather}
F(\mathbf{u},\mathbf{v},\gamma,t)=\delta[\mathbf{u}-\mathbf{u}(t)]
\delta[\mathbf{v}-\mathbf{v}(t)]\delta[\gamma-\gamma(t)],
\label{microscopic}
\\
F(\mathbf{u},\mathbf{v},\gamma,0)=\delta(\mathbf{u})\delta(\mathbf{v})\delta(\gamma).
\label{initialF}
\end{gather}
The joint probability density function is defined as
\begin{equation}
p(\mathbf{u},\mathbf{v},\gamma,t)=\langle
F(\mathbf{u},\mathbf{v},\gamma,t)\rangle.
\end{equation}
Differentiating Eq. (\ref{microscopic}) with respect to $t$, we
have the Liouville equation for $F$,
\begin{equation}
\label{Liouville} \left[\frac{\partial}{\partial
t}+\mathbf{v}\frac{\partial}{\partial
\mathbf{u}}+\frac{\mathbf{f}}{2}\frac{\partial}{\partial
\mathbf{v}}
+\left(\frac{\mathbf{v}^{2}}{4}+\frac{\mathbf{u}\cdot\mathbf{f}}{2}\right)
\frac{\partial}{\partial\gamma}
\right]F(\mathbf{u},\mathbf{v},\gamma,t)=0,
\end{equation}
with the initial condition (\ref{initialF}). Averaging of Eq.
(\ref{Liouville}) gives
\begin{equation}
\label{FFF} \left(\frac{\partial}{\partial
t}+\mathbf{v}\frac{\partial}{\partial
\mathbf{u}}+\frac{\mathbf{v}^{2}}{4}\frac{\partial}{\partial\gamma}\right)\langle
F\rangle+\frac{1}{2}\left(\frac{\partial}{\partial\mathbf{v}}
+\mathbf{u}\frac{\partial}{\partial\gamma}\right)\langle\mathbf{f}F\rangle=0.
\end{equation}
To determine $\langle\mathbf{f}F\rangle$, we use
Donsker-Furutsu-Novikov formula (\ref{Furutsu}),
\begin{equation}
\langle\mathbf{f}(t)F(\mathbf{u},\mathbf{v},\gamma,t)\rangle=\int_{0}^{\infty}
\langle f_{i}(t)f_{j}(t^{\prime})\rangle\left\langle \frac{\delta
F(\mathbf{u},\mathbf{v},\gamma,t)}{\delta
f_{j}(t^{\prime})}\right\rangle dt^{\prime}.
\end{equation}
The functional derivative $\langle\delta F/\delta f_{j}\rangle$
can be found from Eq. (\ref{Liouville}), and then using Eq.
(\ref{correlator2}) one can obtain
\begin{equation}
\langle\mathbf{f}(t)F(\mathbf{u},\mathbf{v},\gamma,t)\rangle=-\frac{3B}{2}
\left(\frac{\partial}{\partial\mathbf{v}}+\mathbf{u}\frac{\partial}{\partial\gamma}\right)
\langle F(\mathbf{u},\mathbf{v},\gamma,t)\rangle.
\end{equation}
Inserting this into Eq. (\ref{FFF}), we finally get the
Fokker-Planck equation,
\begin{gather}
\left[\frac{\partial}{\partial
t}+\mathbf{v}\frac{\partial}{\partial \mathbf{u}}+
\frac{\mathbf{v}^{2}}{4}\frac{\partial}{\partial\gamma}-\frac{3B}{4}
\left(\frac{\partial^{2}}{\partial\mathbf{v}^{2}}+2\mathbf{u}\frac{\partial^{2}}{\partial
\mathbf{v}\partial\gamma} \right. \right. \nonumber
\\
\left.\left.
 +\mathbf{u}^{2}\frac{\partial^{2}}{\partial \gamma^{2}}\right)
\right]p(\mathbf{u},\mathbf{v},\gamma,t)=0, \label{Fokker-general}
\end{gather}
with the initial condition
$p(\mathbf{u},\mathbf{v},\gamma,0)=\delta(\mathbf{u})\delta(\mathbf{v})\delta(\gamma)$.


\begin{thebibliography}{59}

\bibitem{Bass1988}
  F.~G. Bass, Y.~S. Kivshar, V.~V. Konotop,  and Y.~A. Sinitsyn,
  Dynamics of solitons under random perturbations,
  Phys. Rep. \textbf{157}, 63-181 (1988).

\bibitem{Gredeskul1992}
S.~A. Gredeskul and Y.~S. Kivshar, Propagation and   scattering of
nonlinear waves in disordered systems, Phys. Rep. \textbf{216},
1-61 (1992).

\bibitem{Konotop_book1994}
V.~V.Konotop and L.~V\'{a}zquez, \emph{Nonlinear random
  waves} (World Scientific Publishing, Singapore, 1994).

\bibitem {Abdullaev_Garnier2005}
F.~Abdullaev and J.~ Garnier, Optical
  solitons in random media, in \emph{Progress in Optics Vol. 48}, edited by
  E.~Wolf (Elsevier, New York, 2005) pp. 35-106.

\bibitem{Tatarski_book1967}
V.~Tatarskii, \emph{Wave propagation in a Turbulent Medium}
(Dover, New York, 1967).

\bibitem{Ishimaru_book}
A.~Ishimaru, \emph{Wave propagation and Scattering in Random
Media}  (Academic Press, New York, 1978).

\bibitem{Gredeskul1988}
I.~M. Lifshits, S.~A. Gredeskul, and L.~A. Pastur,
\emph{Introduction to the Theory of Disorder Systems} (JohnWiley
\& Sons, New York, 1988).

\bibitem{Abdullaev1982}
F.~Kh. Abdullaev, Soliton collective excitations in stochastic
molecular crystals, Theor. Math. Phys. \textbf{51}, 607-610
(1982).

\bibitem{Elgin1985}
J. N. Elgin, Inverse scattering theory with stochastic initial
potentials, Phys. Lett. A \textbf{110}, 441-443 (1985).

\bibitem{Gordon-Haus1986}
J. P. Gordon and H. A. Haus, Random walk of coherently amplified
solitons in optical fiber transmission, Opt. Lett. \textbf{11},
665-667 (1986).

\bibitem{Malomed1987}
B. A. Malomed, Emission from, quasi-classical quantization, and
stochastic decay of sine-Gordon solitons in external fields,
Physica D \textbf{27}, 113-157  (1987).

\bibitem{Bass-Kivshar1988}
F.~G. Bass, Yu.~S. Kivshar, V.~V. Konotop and S.~A. Puzenko,
Propagation of nonlinear incoherent pulses in single-mode optical
fibres, Opt. Commun. \textbf{68}, 385-390 (1988).

\bibitem{Sanchez1990}
Yu. S. Kivshar, S. A. Gredeskul, A. S\'{a}nchez and L.
V\'{a}zquez, Localization decay induced by strong nonlinearity in
disordered systems, Phys. Rev. Lett. \textbf{64}, 1693-1696
(1990).

\bibitem{Garnier2001}
J.~Garnier, Solitons in random media with long-range correlation,
Waves Random Media \textbf{11}, 149-162 (2001).

\bibitem{Abdulaev2004}
F.~Kh. Abdullaev, D.~V. Navotny, B.~B. Baizakov, Optical pulse
propagation in fibers with random dispersion , Physica D
\textbf{192}, 83-94 (2004).

\bibitem{Conti2012}
C.~Conti, Solitonization of the Anderson localization,
 Phys. Rev. A  \textbf{86}, 061801(R) (2012).

\bibitem{Fishman2012}
E.~Michaely and S.~Fishman, Effective noise
  theory for the nonlinear Schr{\"{o}}dinger equation with disorder,
  Phys. Rev. E  \textbf{85}, 046218 (2012).

\bibitem{Schwiete2013}
G.~Schwiete and A.~M. Finkel'stein, Effective
  theory for the propagation of a wave packet in a disordered and nonlinear
  medium, Phys.  Rev. A \textbf{87}, 043636  (2013).

\bibitem{Weiss2016}
C.~Weiss, S.~L.Cornish, S.~A.Gardiner,  and   H.-P.Breuer,
Superballistic center-of-mass motion in
  one-dimensional attractive bose gases: Decoherence-induced gaussian random
  walks in velocity space, Phys. Rev. A \textbf{93}, 013605 (2016).

\bibitem{Kivshar-Malomed1989}
Yu. S. Kivshar and B. A. Malomed, Dynamics of solitons in nearly
integrable systems, Rev. Mod. Phys. \textbf{61}, 763-915 (1989).

\bibitem{Pikovsky2008}
A. S. Pikovsky and D. L. Shepelyansky, Destruction of Anderson
localization by a weak nonlinearity, Phys. Rev. Lett.
\textbf{100}, 094101 (2008).

\bibitem{Flach2009}
S. Flach, D. O. Krimer, and Ch. Skokos, Universal spreading of
wave packets in disordered nonlinear systems, Phys. Rev. Lett.
\textbf{102}, 024101 (2009).

\bibitem{Milovanov2012}
A. V. Milovanov and A. Iomin, Localization-delocalization
transition on a separatrix system of nonlinear Schr\"{o}dinger
equation with disorder, Europhys. Lett. \textbf{100}, 10006
(2012).

\bibitem{Debussche1999}
A.~Debussche and J.~Printems, Numerical simulations of the
stochastic Korteweg de Vries equation, Physica D \textbf{134},
200-226 (1999).

\bibitem{Debussche2002}
A.~Debussche and L.~D. Menza, Numerical simulation of focusing
stochastic nonlinear Schr\"{o}dinger equations, Physica D
\textbf{162}, 131-154 (2002).

\bibitem{Molli2010}
V.~Folli and C.~Conti, Frustrated brownian motion
  of nonlocal solitary waves, Phys. Rev. Lett. \textbf{104}, 193901
  (2010).

\bibitem{Skupin2012}
F.~Maucher, W.~Krolikowski, and S.~Skupin, Stability of
  solitary waves in random nonlocal nonlinear media,
Phys. Rev. A \textbf{85}, 063803 (2012).

\bibitem{Ruderman2002}
M.~S. Ruderman, Propagation of
  solitons of the derivative nonlinear Schr\"{o}dinger equation in a plasma
  with fluctuating density,  Phys. Plasmas \textbf{9},
  2940-2945 (2002).

\bibitem{Lashkin_Alfen2006}
 V.~M. Lashkin, Alfv\'{e}n soliton and emitted radiation in the presence of perturbations,
 Phys. Rev. E \textbf{74}, 016603 (2006).

  \bibitem{Lashkin2021}
V.~M. Lashkin, Perturbation theory for solitons of the
Fokas-Lenells equation: Inverse scattering
  transform approach, Phys. Rev. E \textbf{103},
 042203 (2021).

\bibitem{Doktorov2001}
 E~ V. Doktorov and I.~S. Kuten, The Gordon-Haus effect for modified NLS solitons,
 Europhys. Lett. \textbf{53}, 22 (2001).

\bibitem{Lashkin_MNLS2004}
 V.~M. Lashkin, Soliton of modified nonlinear Schr\"{o}dinger equation with random perturbations,
 Phys. Rev. E \textbf{69}, 016611 (2004).

\bibitem{Gorodtsov2000}
V.~A. Gorodtsov, Stochastic Kadomtsev--Petviashvili equation, JETP
\textbf{90}, 1105-1113 (2000).

\bibitem{Chiao1964}
 R.~Y. Chiao, E. Garmire, and C.~H. Townes, Self-trapping of optical
  beams, Phys. Rev. Lett. \textbf{13}, 479--482 (1964).

\bibitem{Kelley1965}
P.~L. Kelley, Self-focusing of optical beams, Phys. Rev. Lett.
\textbf{15}, 1005-1008 (1965).

\bibitem{Rubenchik1986}
E.~A. Kuznetsov, A.~M. Rubenchik,  and V.~E. Zakharov, Soliton
stability in plasmas and hydrodynamics, Phys. Rep. \textbf{142},
103-165 (1986).

\bibitem{Zakharov1972}
V.~E. Zakharov, Collapse of Langmuir waves, Sov. Phys. JETP
\textbf{35}, 908-914 (1972).

\bibitem{Zakharov_UFN2012}
V.~E. Zakharov and E.~A. Kuznetsov, Solitons and collapses: two
evolution scenarios of nonlinear wave systems, Physics-Uspekhi
\textbf{55}, 535-556 (2012).

\bibitem{Fibich_book}
G.~Fibich, \emph{The Nonlinear  Schr\"{o}dinger Equation: Singular
Solutions and Optical Collapse}  (Springer, Heidelberg, 2015).

\bibitem{Berge1998}
L.~Berg\'{e}, Wave collapse in physics: principles and
applications to light and plasma waves, Phys. Rep. \textbf{303},
259-370 (1998).

\bibitem{Pasmanik1974}
G.~A. Pasmanik, Self-interaction of incoherent light beams, Sov.
Phys. JETP \textbf{39}, 234-238 (1974).

\bibitem{Jokipii1973}
 J.~R. Jokipii and J. Marburger, Homogeneity requirements for minimizing self-focusing
  damage by strong electromagnetic waves, Appl. Phys. Lett. \textbf{23}, 696-698
  (1973).

\bibitem{Vorobev1970}
V.~V. Vorob'ev, The broadening of a light beam in a nonlinear
medium having random inhomogeneities of the refractive index, Izv.
Vuz. Radiofiz. \textbf{13}, 1053-1057 (1970).

\bibitem{Talanov1971}
S.~N. Vlasov, V.~A. Petrishchev,  and V.~I. Talanov, Averaged
  description of wave beams in linear and nonlinear medis (the method of
  moments), Radiophys. Quantum Electron. \textbf{14}, 1062-1070
  (1974).

\bibitem{Gaididei1998}
Yu.~B. Gaididei and P.~L. Christiansen, Spatiotemporal collapse in
a nonlinear waveguide with a randomly fluctuating refractive
index, Opt. Lett. \textbf{23}, 1090-1092 (1998).

\bibitem{Shivamoggi2000}
B.~K. Shivamoggi, L.~C. Andrews,  and R.~L. Phillips, Nonlinear
wave propagation in a random medium, Physica A \textbf{275}, 86-98
(2000).

\bibitem{Gorder2010}
 R.~A. Van Gorder, $\delta$-expansion method for nonlinear stochastic differential equations
  describing wave propagation in a random medium, Phys. Rev. E \textbf{82}, 056712
  (2010).

\bibitem{Hafizi2017}
 B.~Hafizi, J.~R. Pe\~{n}ano, J.~P. Palastro, R.~P. Fischer, and G.~DiComo, Laser beam
  self-focusing in turbulent dissipative media, Opt. Lett. \textbf{42}, 298-301
  (2017).

 \bibitem{Penano2017}
 J.~Pe\~{n}ano, J.~P. Palastro, B.~Hafizi,
 M.~H. Helle,  and G.~P. DiComo, Self-channeling of
  high-power laser pulses through strong atmospheric turbulence, Phys. Rev. A \textbf{96},
 013829 (2017).

\bibitem{Deng2020}
Y.~Deng, H.~Wang, X.~Ji, X.~Li, H.~Yu, and L.~Chen,
Characteristics of high-power partially coherent laser beams
propagating upwards in the turbulent atmosphere, Opt. Express
\textbf{28}, 27927-27939 (2020).

\bibitem{Garnier2004}
J. Garnier, F. Kh. Abdullaev, and B. B. Baizakov, Collapse of a
Bose-Einstein condensate induced by fluctuations of the laser
intensity, Phys. Rev. A \textbf{69}, 053607 (2004).

\bibitem{Besieris_main}
I.~M. Besieris, Solitons in randomly inhomogeneous media, in
\emph{Nonlinear Electromagnetics}, edited by P.~L.~E. Uslenghi
  (Academic Press, New York, 1980)  pp. 87-116.

\bibitem{Berge2002}
T.~J. Alexander and L.~Berg\'{e}, Ground states and vortices of
matter-wave condensates and optical guided waves, Phys. Rev. E
\textbf{65}, 026611 (2002).

\bibitem{Kivshar_book2003}
Y.~S. Kivshar and G.~P. Agrawal, \emph{Optical Solitons: From
  Fibers to Photonic Crystals} (Academic Press, San Diego, 2003).

\bibitem{Gurevich_book1978}
A.~V. Gurevich, \emph{Nonlinear Phenomena in the Ionosphere}
(Springer-Verlag, Berlin, 1978).

\bibitem{Goldman1984}
M.~V. Goldman, Strong turbulence of plasma waves, Rev. Mod. Phys.
\textbf{56}, 709-735 (1984).

\bibitem{Yukalov2007}
 V.~I. Yukalov, E.~P. Yukalova, K.~V. Krutitsky, and R.~Graham,
Bose-Einstein-condensed gases in arbitrarily strong random
potentials, Phys. Rev. A  \textbf{76}, 053623 (2007).

\bibitem{Yu2017}
Z.-Y. Sun and X.~Yu, Transport of nonautonomous solitons in
two-dimensional disordered media, Ann. Phys., 1600323  (2017).

\bibitem{Furutsu1963}
K.~Furutsu, On the statistical theory of electromagnetic waves in
a fluctuating medium (i), J. Res. NBS \textbf{67D}, 303-323
(1963).

\bibitem{Donsker1964}
M.~D. Donsker, On function space integrals, in \emph{Analysis
  in Function Space}, edited by W.~T. Martin and I.~Segal
   (M.I.T. Press, Cambridge, 1964) pp. 17-30.

\bibitem{Novikov1965}
E.~A. Novikov, Functionals and the random-force method in
turbulence theory, Sov. Phys. JETP \textbf{20}, 1290-1294 (1965).

\bibitem{Jokipii1975}
L.~C. Lee and J.~R. Jokipii, Strong scintillations in
astrophysics. II. A theory of temporal broadening of pulses,
Astrophys. J. \textbf{201}, 532-543 (1975).

\bibitem{Besieris1979}
C.~M. Rose and I.~M. Besieris, $N$th-order multifrequency
coherence functions: A functional path integral approach, J. Math.
Phys. \textbf{20}, 1530-1538 (1979).

\bibitem{Bezak1998}
V.~Bez\'{a}k, Theory of randomly inhomogeneous waveguides with
slight material dispersion: I. Single-mode waveguides, Waves
Random Media \textbf{8}, 351-367 (1998).

\bibitem{Zakharov-book1984}
S.~P. Novikov, S.~V. Manakov, L.~P. Pitaevski, and V.~E. Zakharov,
\emph{Theory of Solitons: The Inverse Scattering Method}
(Consultants Bureau, New York, 1984).

\bibitem{Serkin2001}
V. N. Serkin  and T. L. Belyaeva, High-energy optical
Schr\"{o}dinger solitons, JETP Lett. \textbf{74}, 573-577.
  (2001).

\bibitem{Serkin2002}
V. N. Serkin  and A. Hasegawa, Exactly integrable nonlinear
Schr\"{o}dinger equation models with varying dispersion,
nonlinearity and gain: Application for soliton dispersion
managements, IEEE J. Sel. Top. Quantum Electron. \textbf{8},
418-431  (2002).

\bibitem{Kampen-book1992}
N.~G. van Kampen, \emph{Stochastic Processes in Physics and
Chemistry} (North-Holland, Amsterdam, 1992).

\bibitem{Sulem1999}
C.~Sulem and P.-L. Sulem, \emph{The Nonlinear
  Schr\"{o}dinger Equation} (Springer, New York, 1999).

\bibitem{Lashkin2012}
V.~M. Lashkin, A.~S. Desyatnikov, E.~A. Ostrovskaya, and Yu.~S.
Kivshar, Azimuthal vortex clusters in Bose-Einstein condensates,
Phys. Rev. A \textbf{85}, 013620 (2012).

\bibitem{Malomed_vortex2019}
B.~A. Malomed, Vortex solitons: Old results and new perspectives,
Physica D \textbf{399}, 108-137 (2019).

\bibitem{Louis-Experiment2016}
H.~Louis, M.~Tlidi, and E.~Louvergneaux, Experimental evidence of
dynamical propagation for solitary waves in ultra slow stochastic
non-local Kerr medium, Opt. Express \textbf{24}, 16206-16211
(2016).

\bibitem{Wigner1932}
E.~Wigner, On the quantum correction for thermodynamic
equilibrium, Phys. Rev.  \textbf{40}, 749-759 (1932).

\end{thebibliography}
\end{document}